\documentclass{article}
\usepackage{spconf}
\ninept
\usepackage{amsmath, amssymb, epsfig, graphicx, bm}
\usepackage{cases}
\usepackage{float, subfigure}
\usepackage{color}
\usepackage{amsthm}
\usepackage{amsfonts}
\usepackage{cite, url}
\usepackage[all,cmtip]{xy}
\usepackage{color,hyperref}
\usepackage{algorithmic,algorithm}

\definecolor{darkblue}{rgb}{0,0,1}
\hypersetup{colorlinks,breaklinks,
linkcolor=darkblue,urlcolor=darkblue,anchorcolor=darkblue,citecolor=darkblue}

\usepackage[normalem]{ulem} 

\title{Robust Group LASSO Over Decentralized Networks}

\name{Manxi Wang\textsuperscript{*} \hspace{0.2in} Yongcheng
Li\textsuperscript{*} \hspace{0.2in} Xiaohan
Wei\textsuperscript{\dag} \hspace{0.2in} Qing
Ling\textsuperscript{\ddag}}

\address{
\small \textsuperscript{*} State Key Laboratory of Complex Electromagnetic Environment Effects on Electronics and Information System, Luoyang, China \vspace{-0.03in} \\
\small \textsuperscript{\dag} Department of Electrical Engineering, University of Southern California, Los Angeles, USA \vspace{-0.03in} \\
\small \textsuperscript{\ddag} Department of Automation,
University of Science and Technology of China, Hefei, China
\vspace{-0.03in}}

\begin{document}

\maketitle

\begin{abstract}

This paper considers the recovery of group sparse signals over a
multi-agent network, where the measurements are subject to sparse
errors. We first investigate the robust group LASSO model and its
centralized algorithm based on the alternating direction method of
multipliers (ADMM), which requires a central fusion center to
compute a global row-support detector. To implement it in a
decentralized network environment, we then adopt dynamic average
consensus strategies that enable dynamic tracking of the global
row-support detector. Numerical experiments demonstrate the
effectiveness of the proposed algorithms.

\end{abstract}


\section{Introduction} \label{sec:intro}

Suppose that $L$ distributed agents constitute a bidirectionally
connected network and sense correlated signals under sparse
measurement errors. The measurement equation of agent $l$ is
\begin{equation}\label{ex0}
    \mathbf{m}_l = \mathbf{A}_{(l)}\mathbf{y}_{l} + \mathbf{s}_l,
\end{equation}
where $\mathbf{m}_l \in \mathcal{R}^{M}$ is the measurement
vector, $\mathbf{A}_{(l)}$ is the sensing matrix, $\mathbf{y}_l
\in \mathcal{R}^{N}$ is the unknown signal vector, and
$\mathbf{s}_l \in \mathcal{R}^{M}$ is the unknown sparse error
vector. We are particularly interested in a certain correlation
pattern of the signal vectors, where the signal matrix $\mathbf{Y}
= [\mathbf{y}_1, \ldots, \mathbf{y}_L] \in \mathcal{R}^{N \times
L}$ is group sparse, meaning that $\mathbf{Y}$ is sparse and its
nonzero entries appear in a small number of common rows. Define
$\mathbf{M} \in \mathcal{R}^{M \times L}$ as the measurement
matrix and $\mathbf{S} \in \mathcal{R}^{M \times L}$ as the sparse
error matrix, the matrix form of the agents' measurement equations
is
\begin{equation}\label{ex1}
    \mathbf{M} = [\mathbf{A}_{(1)}\mathbf{y}_{1}, \cdots, \mathbf{A}_{(L)}\mathbf{y}_{L}] +
    \mathbf{S}.
\end{equation}
Given $\mathbf{M}$ and $\mathbf{A}_{(i)}$'s, the goal of the
network is to recover $\mathbf{Y}$ and $\mathbf{S}$ from the
linear measurement equation \eqref{ex1}.

\subsection{Robust Group LASSO Model}

The recovery of group sparse (also known as block sparse
\cite{eldar2010} or jointly sparse \cite{eldar2012}) signals finds
a variety of applications such as direction-of-arrival estimation
\cite{malioutov2005,wei2012}, collaborative spectrum sensing
\cite{tian2011, meng2011, giannakis2011} and motion detection
\cite{gao2014}. A well-known model to recover group sparse signals
is group LASSO (least absolute shrinkage and selection operator)
\cite{yuan2007}, which solves
\begin{align}\label{ee2gl}
  \min\limits_{\mathbf{Y}} \quad \|\mathbf{Y}\|_{2,1} + \lambda \|\mathbf{M} - [\mathbf{A}_{(1)}\mathbf{y}_{1}, \cdots,
  \mathbf{A}_{(L)}\mathbf{y}_{L}]\|_F^2.
\end{align}
Here $\lambda$ is a nonnegative trade-off parameter. A key assumption
leading to the success of such model is the sub-Gaussianity of errors.
However, in many applications,
the measurements of the agents may be seriously contaminated or
even missing due to uncertainties such as sensor failure or
transmission errors. This kind of measurement errors are often
sparse \cite{giannakis2012}. Hence, a natural extension of \eqref{ee2gl} is
to exploit the structures of both the signal
matrix $\mathbf{Y}$ and the sparse error matrix $\mathbf{S}$ by
solving
\begin{align}\label{ee3}
  \min\limits_{\mathbf{Y}, \mathbf{S}} & \quad \|\mathbf{Y}\|_{2,1} + \lambda \|\mathbf{S}\|_{1}, \\
  s.t.                                 & \quad \mathbf{M} = [\mathbf{A}_{(1)}\mathbf{y}_{1},
  \cdots, \mathbf{A}_{(L)}\mathbf{y}_{L}] + \mathbf{S}. \nonumber
\end{align}
This model is termed as
robust group LASSO, whose performance guarantee is given in
\cite{wei2015}. Under mild conditions, the robust group LASSO
model is able to simultaneously recover the true values of
$\mathbf{Y}$ and $\mathbf{S}$ with high probability.

\vspace{-0.7em}

\subsection{Our Contributions}

This paper develops efficient algorithms to solve
the robust group LASSO model \eqref{ee3}. Our contributions are as
follows.

\begin{enumerate}

\item[(i)] We propose a centralized algorithm that
is based on the alternating direction method of multipliers
(ADMM), a powerful operator-splitting technique. One subproblem of
the centralized algorithm is the traditional group LASSO model,
which is approximately solved by a block coordinate descent (BCD)
approach through successively estimating the row-support of the
signal matrix $\mathbf{Y}$.

\item[(ii)] We develop decentralized versions of the above
algorithm that are suitable for autonomous computation over
large-scale networks. Since estimating the row-support of
the signal matrix $\mathbf{Y}$ requires collaborative information
fusion of all the agents, we propose to achieve inexact
information fusion through dynamic average consensus techniques,
which only require information exchange among neighboring agents.
\end{enumerate}

\vspace{-1em}

\subsection{Notations}

Matrices are denoted by bold uppercase letters and vectors are
denoted by bold lowercase letters. For a matrix $\mathbf{D}$,
$\mathbf{d}^i$ denotes its $i$-th row, $\mathbf{d}_j$ denotes its
$j$-th column, while $d_{ij}$ denotes its $(i,j)$-th element. The
$\ell_{2,1}$-norm of $\mathbf{D}$ is $\|\mathbf{D}\|_{2,1}
\triangleq \sum_i (\sum_j d_{ij}^2)^{1/2}$, the $\ell_1$-norm is
$\|\mathbf{D}\|_1 \triangleq \sum_i \sum_j |d_{ij}|$, and the
Frobenius norm is $\|\mathbf{D}\|_F \triangleq (\sum_i \sum_j
d_{ij}^2)^{1/2}$.

The multi-agent network is described as a bidirectional graph
$(\mathcal{L},\mathcal{E})$. If two agents $r, l \in \mathcal{L}$
are neighbors, then they can communicate with each other within
one hop, and $(r,l) \in \mathcal{E}$ is a bidirectional
communication edge.

\section{Centralized Robust Group LASSO} \label{sec:centralized}

Optimally solving (\ref{ee3}) is
nontrivial since the objective function is a weighted summation
of two nonsmooth functions $\|\mathbf{Y}\|_{2,1}$ and
$\|\mathbf{S}\|_1$, where $\mathbf{Y}$ and $\mathbf{S}$ are
entangled in the constraint. Therefore we resort to the
alternating direction method of multipliers (ADMM) to split the
two entangled variables $\mathbf{Y}$ and $\mathbf{S}$ such that
the resulting subproblems are easier to solve.

\subsection{Using ADMM to Solve (\ref{ee3})}

The augmented Lagrangian function of (\ref{ee3}) is
\begin{align}
    \|\mathbf{Y}\|_{2,1}+\lambda\|\mathbf{S}\|_{1} & - \langle \mathbf{Z}, [\mathbf{A}_{(1)}\mathbf{y}_{1}, \cdots, \mathbf{A}_{(L)}\mathbf{y}_{L}]+\mathbf{S}-\mathbf{M} \rangle \nonumber \\
    & +\frac{\beta}{2}\|[\mathbf{A}_{(1)}\mathbf{y}_{1}, \cdots, \mathbf{A}_{(L)}\mathbf{y}_{L}]+\mathbf{S}-\mathbf{M}\|_{F}^{2},
    \nonumber
\end{align}
where $\mathbf{Z}\in \mathcal{R}^{M\times L}$ is the Lagrange
multiplier and $\beta$ is a positive penalty parameter.
The ADMM
alternatingly minimizes the augmented Lagrangian function with
respect to $\mathbf{Y}$ and $\mathbf{S}$, and then updates the
Lagrange multiplier $\mathbf{Z}$ \cite{Bertsekas1999}. At time
$t$, the ADMM works as follows.

First, fixing $\mathbf{S}=\mathbf{S}(t)$ and
$\mathbf{Z}=\mathbf{Z}(t)$, we minimize the augmented Lagrangian
function respect to $\mathbf{Y}$ to get $\mathbf{Y}(t+1)$. Simple
manipulation shows that it is equivalent to
\begin{align}\label{ee12}
\mathbf{Y}(t+1) & = \arg\min_{\mathbf{Y}} \|\mathbf{Y}\|_{2,1} \\
&    +\frac{\beta}{2}\|[\mathbf{A}_{(1)}\mathbf{y}_{1}, \cdots,
\mathbf{A}_{(L)}\mathbf{y}_{L}]+\mathbf{S}(t)-\mathbf{M}-\frac{\mathbf{Z}(t)}{\beta}\|_{F}^{2}.
\nonumber
\end{align}
Note that (\ref{ee12}) is a standard group lasso problem that
generally does not have a closed-form solution. We will develop an
efficient algorithm to solve (\ref{ee12}) later in this section.

Second, fixing $\mathbf{Y}=\mathbf{Y}(t+1)$ and
$\mathbf{Z}=\mathbf{Z}(t)$, we minimize the augmented Lagrangian
function respect to $\mathbf{S}$ to get $\mathbf{S}(t+1)$. Again,
combining the linear term with the quadratic term of $\mathbf{S}$
yields
\begin{align}\label{ee10}
    & \mathbf{S}(t+1) =\textrm{arg}\min_{\mathbf{S}} \lambda\|\mathbf{S}\|_{1} \\
    &  +\frac{\beta}{2}\|[\mathbf{A}_{(1)}\mathbf{y}_{1}(t+1), \cdots,
    \mathbf{A}_{(L)}\mathbf{y}_{L}(t+1)]+\mathbf{S}-\mathbf{M}-\frac{\mathbf{Z}(t)}{\beta}\|_{F}^{2}.
    \nonumber
\end{align}
Denoting
$\mathbf{W}(t+1)=\mathbf{M}-[\mathbf{A}_{(1)}\mathbf{y}_{1}(t+1),
\cdots, \mathbf{A}_{(L)}\mathbf{y}_{L}(t+1)]-\mathbf{Z}(t)/\beta$,
(\ref{ee10}) has a closed-form solution given by
\begin{equation}\label{ee11}
    s_{ml}(t+1)=\textrm{sgn}(w_{ml}(t+1))\max\big(0,|w_{ml}(t+1)|-\frac{\lambda}{\beta}\big),
\end{equation}
where $\textrm{sgn}(\cdot)$ is the sign function; $s_{ml}(t+1)$
and $w_{ml}(t+1)$ denote the $(m,l)$-th entries of
$\mathbf{S}(t+1)$ and $\mathbf{W}(t+1)$, respectively. Note that
the term $|s_{ml}(t+1)|$ can be viewed as the support detector of
the $(m,l)$-th element of $\mathbf{S}$. If $|s_{ml}(t+1)|$ is
smaller than the threshold $\lambda/\beta$, then $s_{ml}(t+1)$ is
set to be zero.

Finally, given $\mathbf{Y}=\mathbf{Y}(t+1)$ and
$\mathbf{S}=\mathbf{S}(t+1)$, the Lagrange multiplier $\mathbf{Z}$
is updated according to the following formula
\begin{align}\label{ee19}
&     \mathbf{Z}(t+1)=\mathbf{Z}(t) \\
&    -\beta\big([\mathbf{A}_{(1)}\mathbf{y}_{1}(t+1), \cdots,
    \mathbf{A}_{(L)}\mathbf{y}_{L}(t+1)]+\mathbf{S}(t+1)-\mathbf{M}\big).
    \nonumber
\end{align}

Since the update of $\mathbf{S}$ in (\ref{ee11}) and the update of
$\mathbf{Z}$ in (\ref{ee19}) are both simple, now we focus on the
update of $\mathbf{Y}$ in (\ref{ee12}) that is the bottleneck of
the ADMM. Observe that in (\ref{ee12}) the $\ell_{2,1}$-norm term
is separable with respect to $\mathbf{y}_i$'s but nonsmooth, while
the Frobenius term is smooth but nonseparable with respect to
$\mathbf{y}_i$'s. Therefore, in this paper we solve (\ref{ee12})
with the block coordinate descent (BCD) algorithm that has shown
to be an efficient tool to handle this special problem structure
\cite{tseng2009,yang2009,luo2012}.

\subsection{Using BCD to Solve (\ref{ee12})}

To set up the iterative BCD algorithm that solves (\ref{ee12}) at
time $t$, we divide time $t$ into $P$ slots. At time $t$ slot $p$
($p=0, 1, \cdots, P-1$), we linearize the Frobenius norm term in
(\ref{ee12}) with respect to $\mathbf{Y}(t+\frac{p}{P})$ and add
an extra quadratic regularization term, which gives
\begin{equation}\label{ee13}
   \min_{\mathbf{Y}}~\|\mathbf{Y}\|_{2,1}+\beta
    \langle \mathbf{V}(t+\frac{p}{P}), \mathbf{Y} \rangle+\frac{\beta}{2\tau}\|\mathbf{Y}-\mathbf{Y}(t+\frac{p}{P})\|_{F}^{2},
\end{equation}
where $\tau$ is a positive proximal parameter and the $l$-th
column of $\mathbf{V}(t+\frac{p}{P})\in \mathcal{R}^{N\times L}$
is defined as
\begin{equation}\label{ee14}
    \mathbf{v}_{l}(t+\frac{p}{P})=\mathbf{A}_{(l)}^{T}\big(\mathbf{A}_{(l)}\mathbf{y}_{l}(t+\frac{p}{P})+\mathbf{s}_{l}(t)-\mathbf{m}_{l}-\frac{\mathbf{z}_{l}(t)}{\beta}\big).
\end{equation}
Note that (\ref{ee13}) is equivalent to
\begin{equation}\label{ee13_tmp}
    \min_{\mathbf{Y}}~\|\mathbf{Y}\|_{2,1}+\frac{\beta}{2\tau}\|\mathbf{Y}-\mathbf{Y}(t+\frac{p}{P})+ \tau \mathbf{V}(t+\frac{p}{P})\|_{F}^{2},
\end{equation}
which has a closed-form solution given by the soft-thresholding
operator \cite{beck2009}. Denote $\mathbf{U}(t+\frac{p}{P}) =
\mathbf{Y}(t+\frac{p}{P})- \tau \mathbf{V}(t+\frac{p}{P}) \in
\mathcal{R}^{N\times L}$ whose $n$-th row is given by
$\mathbf{u}^{n}(t+\frac{p}{P}) = \mathbf{y}^{n}(t+\frac{p}{P})
-\tau \mathbf{v}^{n}(t+\frac{p}{P})$. Also denote
$\mathbf{Y}(t+\frac{p+1}{P}) \in \mathcal{R}^{N\times L}$ as the
solution of (\ref{ee13_tmp}). The $n$-th row of
$\mathbf{Y}(t+\frac{p+1}{P})$ is
\begin{align}\label{ee15}
 \mathbf{y}^{n}(t+\frac{p+1}{P})
=\frac{\mathbf{u}^{n}(t+\frac{p}{P})}{\|\mathbf{u}^{n}(t+\frac{p}{P})\|_{2}}
    \max\big(0,\|\mathbf{u}^{n}(t+\frac{p}{P})\|_{2}-\frac{\tau}{\beta}\big).
    \nonumber
\end{align}

Again, note that the term $\|\mathbf{u}^{n}(t+\frac{p}{P})\|_{2}$
can be viewed as the row-support detector of the $n$-th row of
$\mathbf{Y}$. If $\|\mathbf{u}^{n}(t+\frac{p}{P})\|_{2}$ is
smaller than the threshold $\tau/\beta$, then
$\mathbf{y}^{n}(t+\frac{p+1}{P})$ is set to be zero.

\subsection{Implementation of Centralized Robust Group LASSO}

The centralized ADMM to solve the robust group LASSO model
(\ref{ee3}) is summarized in Table I. Each iteration of the ADMM
includes an inner-loop BCD subroutine that updates $\mathbf{Y}$
through solving (\ref{ee12}), the update of $\mathbf{S}$ that has
a closed-form solution (\ref{ee11}), and the update of
$\mathbf{Z}$ in (\ref{ee19}). The ADMM parameter $\beta$ can be
any positive value, though its choice may influence the
convergence rate. The BCD parameter $\tau$ is set to be the
minimum of largest eigenvalues of
$\mathbf{A}_{(l)}^{T}\mathbf{A}_{(l)},~l=1,2,\cdots,L$ that
guarantees the convergence of the BCD subroutine
\cite{tseng2009,yang2009,luo2012}. As long as $\tau$ is properly
chosen and $P$ is large enough, the BCD subroutine is able to
solve the subproblem (\ref{ee12}) with enough accuracy such that
the ADMM converges to the global minimum of the convex program
(\ref{ee3}).

The algorithm outlined in Table I is centralized, which means that
a fusion center is necessary to gather information from all the
agents and conduct optimization. This centralized scheme is
sensitive to the failure of the fusion center, requires multi-hop
communication within the network, and is hence unscalable with
respect to the networks size. In view of the need of decentralized
optimization for large-scale networks, we discuss how to implement
it in a decentralized manner, as shown in the next section.

\normalsize \hspace{-1em}
\begin{table}
\begin{center}\label{tab1}
\caption{Algorithm 1: Centralized Robust Group LASSO}
\begin{tabular}{l}
  \hline
  Given: measurement $\mathbf{M}$; sensing matrices $\mathbf{A}_{(l)}$; parameters $\beta$ and
  $\tau$\\
  Initialize: signal $\mathbf{Y}(0)=\mathbf{0}$; error $\mathbf{S}(0)=\mathbf{0}$; multiplier $\mathbf{Z}(0)=\mathbf{0}$\\
  \textbf{while} not converged ($t=0,1,\cdots$) \textbf{for all} $l$ \textbf{do}\\
  \hspace*{1mm} \textbf{for} $p=0,1,\cdots, P-1$ \\
  \hspace*{2mm}
  $\mathbf{v}_{l}(t+\frac{p}{P})=\mathbf{A}_{(l)}^{T}\big(\mathbf{A}_{(l)}\mathbf{y}_{l}(t+\frac{p}{P})+\mathbf{s}_{l}(t)-\mathbf{m}_{l}-\frac{\mathbf{z}_{l}(t)}{\beta}\big)$ \\
  \hspace*{2mm}
  $u_{nl}(t+\frac{p}{P}) = y_{nl}(t+\frac{p}{P}) - \tau v_{nl}(t+\frac{p}{P})$, $\forall n$ \\
  \hspace*{2mm} $y_{nl}(t+\frac{p+1}{P})=\frac{y_{nl}(t+\frac{p}{P})}{\|\mathbf{u}^{n}(t+\frac{p}{P})\|_{2}}
    \max\big(0,\|\mathbf{u}^{n}(t+\frac{p}{P})\|_{2}-\frac{\tau}{\beta}\big)$, $\forall n$ \\
  \hspace*{1mm} \textbf{end for}\\
  \hspace*{1mm} $\mathbf{w}_l(t+1)=\mathbf{m}_l-\mathbf{A}_{(l)}\mathbf{y}_{l}(t+1)-\frac{\mathbf{z}_l(t)}{\beta}$\\
  \hspace*{1mm} $s_{ml}(t+1)=\textrm{sgn}(w_{ml}(t+1))\max\big(0,|w_{ml}(t+1)|-\frac{\lambda}{\beta}\big)$, $\forall m$\\
  \hspace*{1mm} $\mathbf{z}_l(t+1)=\mathbf{z}_l(t)-\beta\big(\mathbf{A}_{(l)}\mathbf{y}_{l}(t+1)+\mathbf{s}_l(t+1)-\mathbf{m}_l\big)$\\
  \textbf{end while}\\
  \hline
\end{tabular}
\end{center}
\end{table}

\section{Decentralized Robust Group LASSO} \label{sec:decentralized}

Observe that Algorithm 1 is naturally distributed, except for the
update of $y_{nl}(t+\frac{p+1}{P})$, which involves calculating
the global row-support detector
$\|\mathbf{u}^{n}(t+\frac{p}{P})\|_{2}$ across agents. Hence,
given the vector $\mathbf{u}^{n}(t+\frac{p}{P})$, the key to the
decentralized implementation of Algorithm 1 is how to calculate
its $\ell_2$-norm $\|\mathbf{u}^{n}(t+\frac{p}{P})\|_{2}$ in a
decentralized manner. Recall that
$$\|\mathbf{u}^{n}(t+\frac{p}{P})\|_{2} = L^{\frac{1}{2}} \left(\frac{1}{L} \sum_{l=1}^L u_{nl}^2(t+\frac{p}{P})
\right)^{\frac{1}{2}} = \left( L
h_{nl}(t+\frac{p}{P})\right)^{\frac{1}{2}},$$ where
$$h_{nl}(t+\frac{p}{P}) \triangleq \frac{1}{L}\sum_{l=1}^L
u_{nl}^2(t+\frac{p}{P})$$ is the average of the squares.
Therefore, the problem becomes: Suppose each agent $l$ holds the
value of $u_{nl}^2(t+\frac{p}{P})$, how can we design efficient
strategies to (exactly or inexactly) calculate their mean
$h_{nl}(t+\frac{p}{P})$ in a decentralized manner? Below we
consider three approaches to obtain the average.

\subsection{Static Average Consensus}

The first strategy comes from the classic average consensus
algorithm \cite{boyd2004}. Calculate
$$\mathbf{h}^{n}(t+\frac{p}{P}) = \boldsymbol{\Sigma}^K \left(
\mathbf{u}^{n}(t+\frac{p}{P}) \right)^2,$$ where
$\mathbf{h}^{n}(t+\frac{p}{P}) \in \mathcal{R}^{1 \times L}$ is a
row vector containing all $h_{nl}(t+\frac{p}{P})$, $\left(
\mathbf{u}^{n}(t+\frac{p}{P}) \right)^2$ means element-wise
squares of $\mathbf{u}^{n}(t+\frac{p}{P})$, $K$ is a large
iteration number, and $\boldsymbol{\Sigma}$ is the mixing matrix.
The mixing matrix $\boldsymbol{\Sigma}$ is doubly stochastic, and
its $(r,l)$-th element $\sigma_{rl}$ is nonzero if and only if
$(r,l) \in \mathcal{E}$ or $r=l$. A typical choice of
$\boldsymbol{\Sigma}$ follows the Metropolis-Hastings rule
\cite{boyd2004},
\begin{equation}\label{extra6}
    \sigma_{rl}=\left\{
                   \begin{array}{ll}
                     \min\{\frac{1}{d_{r}},\frac{1}{d_{l}}\}, & \hbox{if $(r,l)\in\mathcal{E}$;} \\
                     \sum_{(r,l)\in\mathcal{E}}\max\left\{0,\frac{1}{d_{r}}-\frac{1}{d_{l}}\right\}, & \hbox{if $r=l$;} \\
                     0, & \hbox{else.}
                   \end{array}
                 \right.
\end{equation}
Here $d_l$ is the degree of agent $l$.

Obviously, the graph-sparse structure of the mixing matrix
$\boldsymbol{\Sigma}$ enables decentralized computation of
$\mathbf{h}^{n}(t+\frac{p}{P})$. According to the theory of
average consensus \cite{boyd2004}, if $K$ goes to infinity, then
all the elements of $\mathbf{h}^{n}(t+\frac{p}{P})$ converge to
the expected average $(1/L)\sum_{l=1}^L u_{nl}^2(t+\frac{p}{P})$,
in which the decentralized implementation is equivalent
to its centralized counterpart. However, increasing $K$
means introducing more rounds of communication and computation,
implying that setting $K$ large is inefficient. On the
other hand, setting $K$ small (say, $K=1$) often
leads to unsatisfactory result.

\subsection{Dynamic Average Consensus}

The above-mentioned dilemma motivates us to introduce a new
scheme to dynamically calculate the
row-support detector. To simplify the algorithmic protocol, we allow
neighboring agents to exchange only one round of information.
Under this setting, every agent holds a dynamic value
$u_{nl}^2(t+\frac{p}{P})$, while all the agents manage to track
their dynamic average with one round of communication. Apparently,
if the values of $u_{nl}^2(t+\frac{p}{P})$ change irregularly, the
agents have no chance to reach their exact dynamic average.
Nevertheless, observe that if the values of
$u_{nl}^2(t+\frac{p}{P})$ converge to their steady states,
convergence of the dynamic average will be possible. We consider
two dynamic average consensus strategies proposed by
\cite{zhu2010}.

\noindent \textbf{First-order dynamic average consensus.}
Calculate
$$h_{nl}(t+\frac{p}{P}) = \sum_{r \neq l} \sigma_{rl}
\left(h_{nr}(t+\frac{p-1}{P}) - h_{nl}(t+\frac{p-1}{P}) \right)$$
$$+ h_{nl}(t+\frac{p-1}{P})  + u_{nl}^2(t+\frac{p}{P}) -
u_{nl}^2(t+\frac{p-1}{P}).$$

\noindent \textbf{Second-order dynamic average consensus.}
Calculate
$$\hspace{-1em} \tilde{h}_{nl}(t+\frac{p}{P}) = u_{nl}^2(t+\frac{p}{P})
- 2 u_{nl}^2(t+\frac{p-1}{P}) + u_{nl}^2(t+\frac{p-2}{P})$$
$$+\tilde{h}_{nl}(t+\frac{p-1}{P})+ \sum_{r \neq l} \sigma_{rl}
\left(\tilde{h}_{nr}(t+\frac{p-1}{P}) -
\tilde{h}_{nl}(t+\frac{p-1}{P}) \right),$$
$$\hspace{-17em} h_{nl}(t+\frac{p}{P}) =
\tilde{h}_{nl}(t+\frac{p}{P})$$ $$+ h_{nl}(t+\frac{p-1}{P})  +
\sum_{r \neq l} \sigma_{rl} \left(h_{nr}(t+\frac{p-1}{P}) -
h_{nl}(t+\frac{p-1}{P}) \right).$$

\normalsize \hspace{-1em}
\begin{table}
\begin{center}\label{tab1}
\caption{Algorithm 2: Decentralized Robust Group LASSO}
\begin{tabular}{l}
  \hline
  Given: measurement $\mathbf{M}$; sensing matrices $\mathbf{A}_{(l)}$; parameters $\beta$ and
  $\tau$\\
  Initialize: signal $\mathbf{Y}(0)=\mathbf{0}$; error $\mathbf{S}(0)=\mathbf{0}$; multiplier $\mathbf{Z}(0)=\mathbf{0}$\\
  \textbf{while} not converged ($t=0,1,\cdots$) \textbf{agent} $l$ \textbf{do}\\
  \hspace*{1mm} \textbf{for} $p=0,1,\cdots, P-1$ \\
  \hspace*{2mm}
  $\mathbf{v}_{l}(t+\frac{p}{P})=\mathbf{A}_{(l)}^{T}\big(\mathbf{A}_{(l)}\mathbf{y}_{l}(t+\frac{p}{P})+\mathbf{s}_{l}(t)-\mathbf{m}_{l}-\frac{\mathbf{z}_{l}(t)}{\beta}\big)$\\
  \hspace*{2mm}
  $u_{nl}(t+\frac{p}{P}) = y_{nl}(t+\frac{p}{P}) - \tau v_{nl}(t+\frac{p}{P})$, $\forall n$ \\
    \hspace*{2mm}
  $h_{nl}(t+\frac{p}{P})$ is updated through an average consensus strategy \\
  \hspace*{2mm} $y_{nl}(t+\frac{p+1}{P})=\frac{y_{nl}(t+\frac{p}{P})}{\sqrt{L h_{nl}(t+\frac{p}{P})}}
    \max\big(0,\sqrt{L h_{nl}(t+\frac{p}{P})}-\frac{\tau}{\beta}\big)$, $\forall n$ \\
  \hspace*{1mm} \textbf{end for}\\
  \hspace*{1mm} $\mathbf{w}_l(t+1)=\mathbf{m}_l-\mathbf{A}_{(l)}\mathbf{y}_{l}(t+1)-\frac{\mathbf{z}_l(t)}{\beta}$, \\
  \hspace*{1mm} $s_{ml}(t+1)=\textrm{sgn}(w_{ml}(t+1))\max\big(0,|w_{ml}(t+1)|-\frac{\lambda}{\beta}\big)$, $\forall m$\\
  \hspace*{1mm} $\mathbf{z}_l(t+1)=\mathbf{z}_l(t)-\beta\big(\mathbf{A}_{(l)}\mathbf{y}_{l}(t+1)+\mathbf{s}_l(t+1)-\mathbf{m}_l\big)$\\
  \textbf{end while}\\
  \hline
\end{tabular}
\end{center}
\end{table}

\setcounter{figure}{1}
\begin{figure*}[ht]
\centering
 \begin{minipage}{5.6cm}
   \includegraphics[height=4cm] {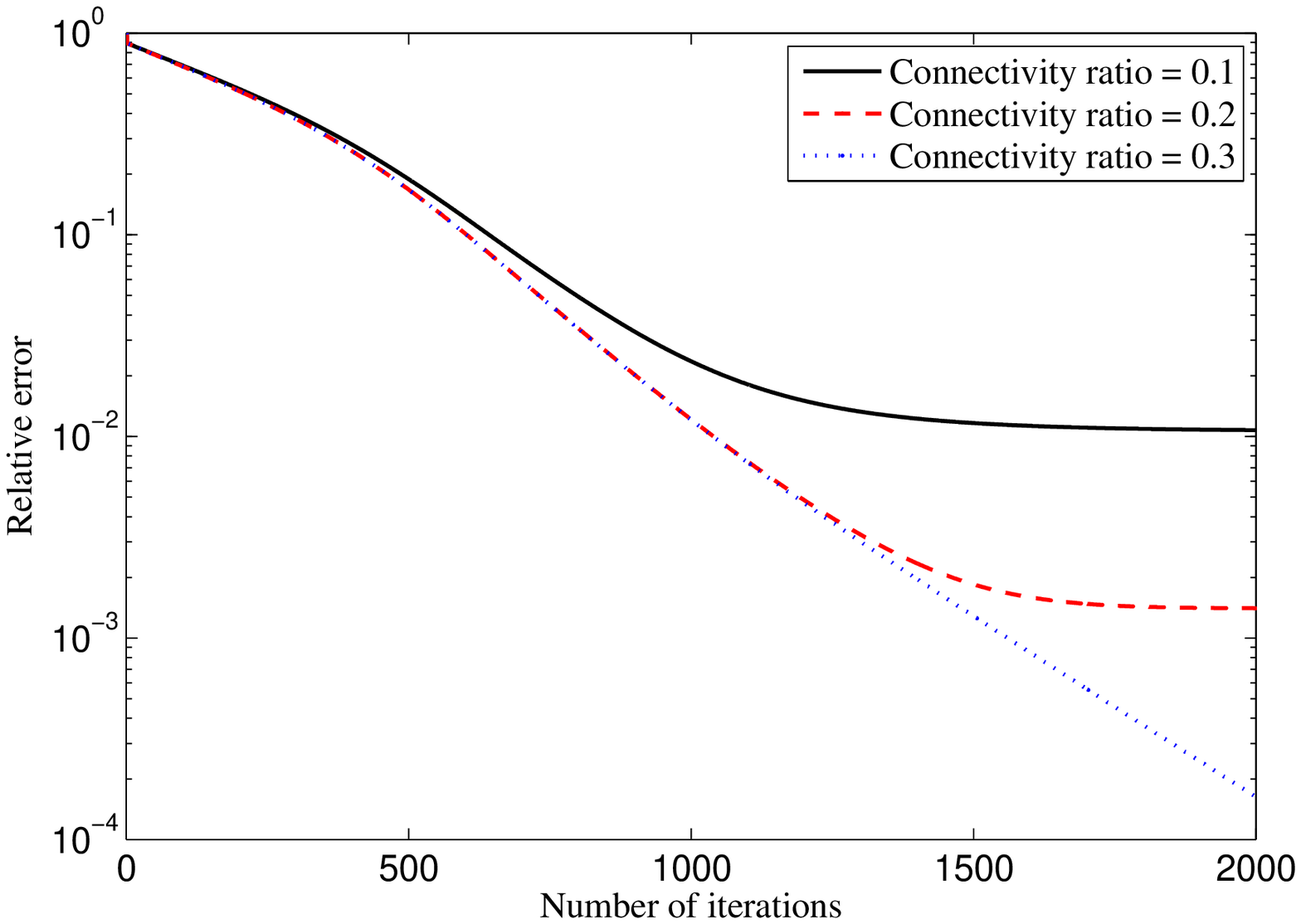}
 \end{minipage}
 \begin{minipage}{5.6cm}
   \includegraphics[height=4cm] {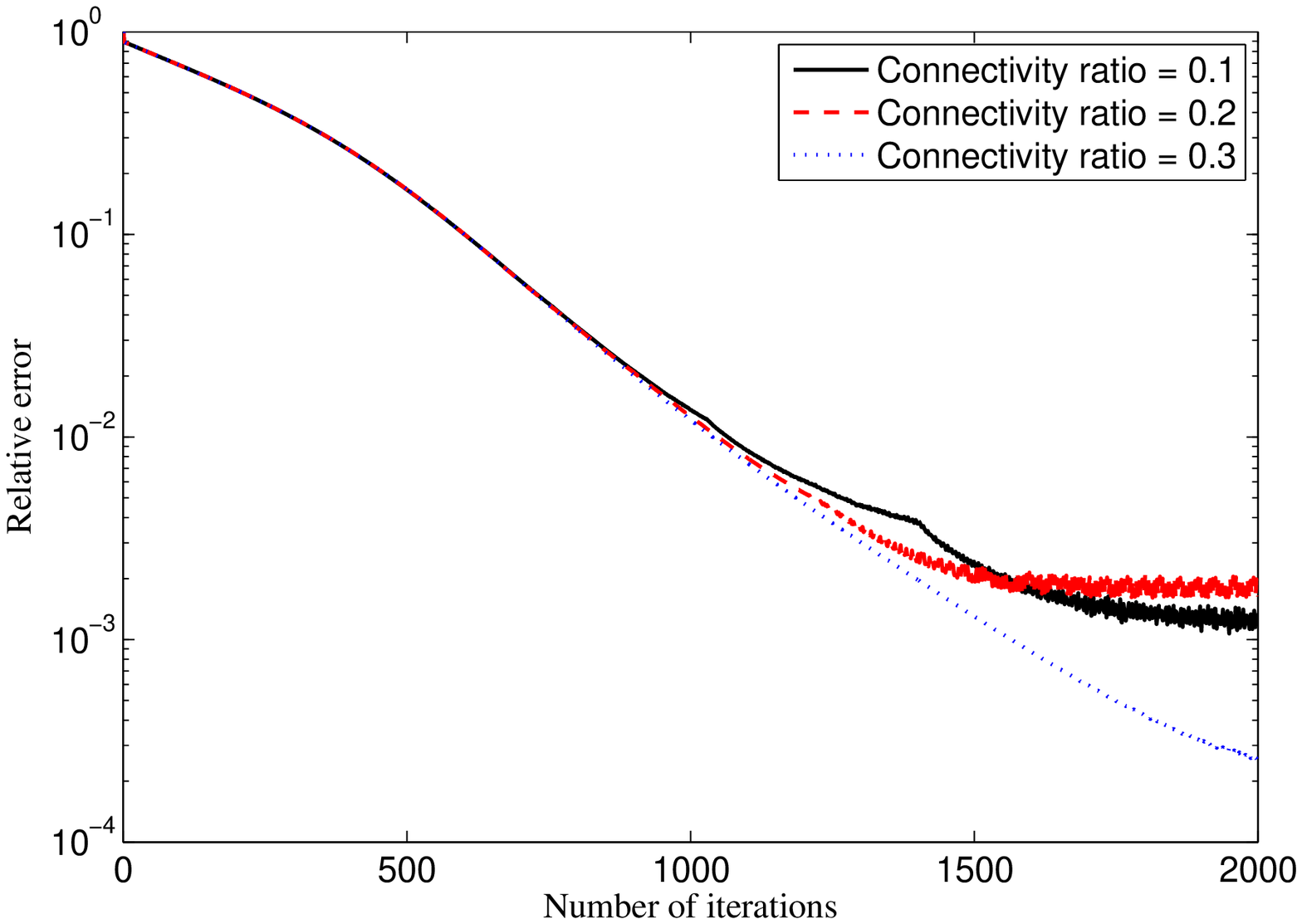}
 \end{minipage}
 \begin{minipage}{5.6cm}
   \includegraphics[height=4cm] {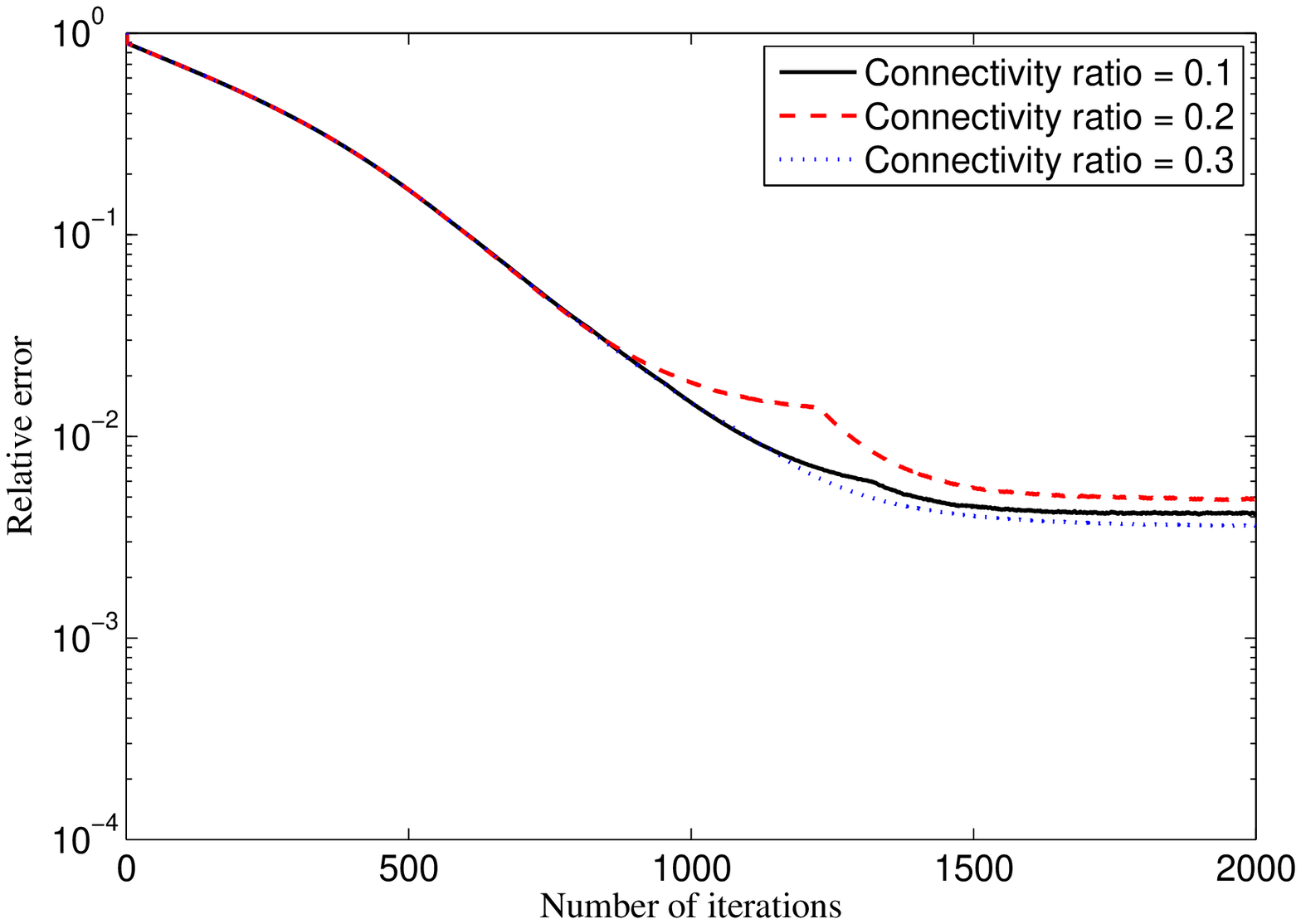}
 \end{minipage}
\caption{Impact of connectivity ratio on the convergence of
decentralized algorithms: static average consensus (Left),
second-order dynamic average consensus (Middle), and first-order
dynamic average consensus (Right).}\label{eps:fig2}
\end{figure*}

\subsection{Implementation of Centralized Robust Group LASSO}

The decentralized group LASSO algorithm is outlined in Table II.
It is very close to the centralized algorithm in Table I, except
that the row-support detector is successively approximated through
static and dynamic average consensus strategies.

If the static average consensus strategy is adopted, then at time
$t$ slot $p$, the network needs $K$ rounds of information
exchange. The number of round reduces to one in the two dynamic
average consensus strategies. Observe that in each round of
first-order dynamic average consensus, agent $l$ requires $h_{nr}$
from all of its neighbors $r$. However, in each round of
second-order dynamic average consensus, agent $l$ requires both
$h_{nr}$ and $\tilde{h}_{nr}$ from all of its neighbors $r$.
Therefore, the second-order strategy doubles the communication
cost per time slot, compared to its first-order counterpart.

With particular note, when $K$ is set to be large enough in the
static average consensus strategy, the average consensus is exact.
Therefore, the resulting decentralized algorithm enjoys the same
convergence guarantee as the centralized one, at the cost of
unaffordable communication cost. Embedding the two dynamic average
consensus strategies saves remarkable communication cost, but
makes convergence analysis a challenging task. We will leave it as
our future work.

In addition, to avoid possible computational instability, we also
set safeguards to the value of $h_{nl}(t+\frac{p}{P})$. If going
beyond the region of $[h_{\min}, h_{\max}]$, its value is set to
the nearest boundary.

\setcounter{figure}{0}
\begin{figure}
\centering \hspace{-2em}\includegraphics[height=6cm]{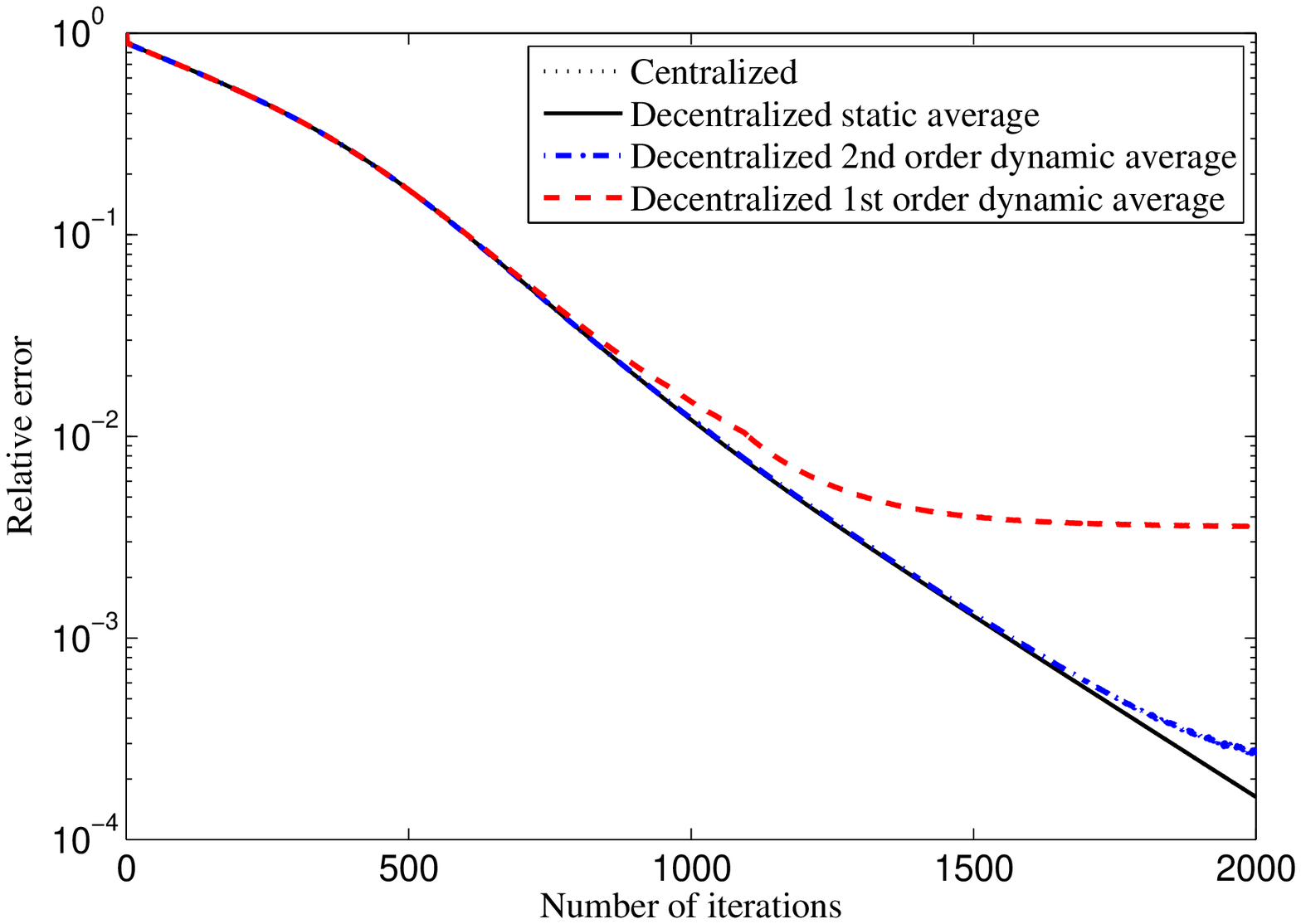}
\caption{Comparison between the centralized algorithm and the
three decentralized ones. The curve of the centralized algorithm
coincides with that using static average
consensus.}\label{eps:fig1}
\end{figure}

\section{Numerical Experiments} \label{sec:simulation}

In the numerical experiments, we consider a network of $L=30$
agents. The dimension of every signal vector is $N=200$, while the
dimension of every measurement vector is $M=30$. The group sparse
signal matrix $\mathbf{Y} \in \mathcal{R}^{200 \times 30}$ has
$10$ nonzero rows (row sparsity ratio is $5\%$), whose positions
are uniformly randomly chosen. The amplitudes of the nonzero
elements follow i.i.d. uniform distribution within $[-50,50]$.
Elements of every sensing matrix $\mathbf{A}_{(l)} \in
\mathcal{R}^{30 \times 200}$ follow i.i.d. standard normal
distribution. The sparse error matrix $\mathbf{S} \in
\mathcal{R}^{30 \times 30}$ has $90$ nonzero elements (sparsity
ratio is $10\%$), whose positions are uniformly randomly chosen
and the amplitudes follow i.i.d. uniform distribution within
$[-50,50]$.

In the robust group LASSO model, the weight parameter $\lambda=1$.
The ADMM parameter $\beta$ is also set as $1$. The BCD parameter
$\tau$ is set to be the minimum of largest eigenvalues of
$\mathbf{A}_{(l)}^{T}\mathbf{A}_{(l)},~l=1,2,\cdots,L$. Every
iteration of the ADMM algorithm is divided into $P=50$ slots so as
to run the BCD subroutine. For the static average consensus
strategy, we let $K=50$, meaning that each slot requires $50$
rounds of communication. For the dynamic average consensus
strategies, we let the safeguards $h_{\min}=1$ and
$h_{\max}=\infty$. The performance metric is relative error,
defined as the Frobenius distance between the true
$[\mathbf{Y}^T~\mathbf{S}^T]$ solving
\eqref{ee3}
and the estimated one by ADMM, normalized by the Frobenius
norm of $[\mathbf{Y}^T~\mathbf{S}^T]$.

We first compare the centralized algorithm and the three
decentralized ones, as depicted in Fig. \ref{eps:fig1}. The
connectivity ratio of the network (the percentage of randomly
connected edges out of all possible ones) is $50\%$. The curve of
the centralized algorithm coincides with that using static average
consensus. Recall that static average consensus incurs $50$ round
of communications at every time slot, and is hence expensive. In
contrast, the dynamic average consensus strategies demonstrate
satisfactory convergence properties, though yielding slightly
degraded estimates. Particularly, the second-order dynamic average
consensus is close to the centralized one in terms of the relative
error.

In the second set of numerical experiments, we vary the
connectivity ratio to observe its impact on the decentralized
algorithms, as shown in Fig. \ref{eps:fig2}. When the connectivity
ratio decreases, the performance of the static average consensus
degrades significantly. The reason is that a lower connectivity
ratio reduces the speed of network information fusion, and hence
makes the static average consensus less accurate under a given
$K$. The two dynamic average consensus strategies, on the other
hand, are not very sensitive to the variation of connectivity
ratio.

The numerical experiments validate the effectiveness of using
dynamic average consensus to decentralize computation over
networks. Though its theoretical properties in tracking problems
have been investigated \cite{zhu2010}, its interplay with the
overall optimization scheme is still unclear, and shall be our
future research focus.

\vspace{0.7em}

\noindent \textbf{Acknowledgement.} Qing Ling is supported in part
by NSF China grant 61573331 and NSF Anhui grant 1608085QF130.

\end{document}